\pdfoutput=1

\documentclass[12pt,a4paper]{article}

\usepackage{ifthen} 
\newboolean{pdflatex}
\setboolean{pdflatex}{true} 

\newboolean{articletitles}
\setboolean{articletitles}{true} 

\newboolean{uprightparticles}
\setboolean{uprightparticles}{false} 

\newboolean{inbibliography}
\setboolean{inbibliography}{false}

\def\paperauthors{LHCb collaboration} 
\def\paperasciititle{} 
\def\papertitle{Searches for the rare hadronic decays \BdPPbarPPbar and \BsPPbarPPbar}

\def\paperkeywords{{High Energy Physics}, {LHCb}} 
\def\papercopyright{CERN on behalf of the LHCb collaboration}

\def\paperlicence{CC BY 4.0 licence}
\def\paperlicenceurl{https://creativecommons.org/licenses/by/4.0/}

\usepackage[top=1in, bottom=1.25in, left=1in, right=1in]{geometry}

\usepackage{microtype}
\usepackage{lineno}  
\usepackage{xspace} 
\usepackage{caption} 

\usepackage{graphicx}  
\usepackage{color}
\usepackage{colortbl}
\graphicspath{{./figs/}} 

\usepackage{amsmath} 
\usepackage{amssymb}
\usepackage{amsfonts}
\usepackage{upgreek} 

\newcommand*\patchAmsMathEnvironmentForLineno[1]{%
\expandafter\let\csname old#1\expandafter\endcsname\csname #1\endcsname
\expandafter\let\csname oldend#1\expandafter\endcsname\csname
end#1\endcsname
 \renewenvironment{#1}%
   {\linenomath\csname old#1\endcsname}%
   {\csname oldend#1\endcsname\endlinenomath}%
}
\newcommand*\patchBothAmsMathEnvironmentsForLineno[1]{%
  \patchAmsMathEnvironmentForLineno{#1}%
  \patchAmsMathEnvironmentForLineno{#1*}%
}
\AtBeginDocument{%
\patchBothAmsMathEnvironmentsForLineno{equation}%
\patchBothAmsMathEnvironmentsForLineno{align}%
\patchBothAmsMathEnvironmentsForLineno{flalign}%
\patchBothAmsMathEnvironmentsForLineno{alignat}%
\patchBothAmsMathEnvironmentsForLineno{gather}%
\patchBothAmsMathEnvironmentsForLineno{multline}%
\patchBothAmsMathEnvironmentsForLineno{eqnarray}%
}

\usepackage{hyperxmp}

\usepackage[pdftex,
            pdfauthor={\paperauthors},
            pdftitle={\paperasciititle},
            pdfkeywords={\paperkeywords},
            pdfcopyright={Copyright (C) \papercopyright},
            pdflicenseurl={\paperlicenceurl}]{hyperref}

\usepackage[colorinlistoftodos,textsize=scriptsize]{todonotes}

\usepackage[bottom,flushmargin,hang,multiple]{footmisc}

\usepackage[all]{hypcap} 

\usepackage{xspace} 
\usepackage{upgreek}

\def\lhcb   {\mbox{LHCb}\xspace}

\def\babar  {\mbox{BaBar}\xspace}

\def\MagUp {\mbox{\em Mag\kern -0.05em Up}\xspace}

\ifthenelse{\boolean{uprightparticles}}%
{

 \def\Ppi         {\ensuremath{\uppi}\xspace}

 \def\Ppsi        {\ensuremath{\uppsi}\xspace}

 \def\PDelta      {\ensuremath{\Delta}\xspace}                 
 \def\PXi         {\ensuremath{\Xi}\xspace}                 
 \def\PLambda     {\ensuremath{\Lambda}\xspace}                 
 \def\PSigma      {\ensuremath{\Sigma}\xspace}                 
 \def\POmega      {\ensuremath{\Omega}\xspace}                 
 \def\PUpsilon    {\ensuremath{\Upsilon}\xspace}

 \def\PB      {\ensuremath{\mathrm{B}}\xspace}                 
                  
 \def\PD      {\ensuremath{\mathrm{D}}\xspace}

 \def\PJ      {\ensuremath{\mathrm{J}}\xspace}                 
 \def\PK      {\ensuremath{\mathrm{K}}\xspace}

 \def\Pb      {\ensuremath{\mathrm{b}}\xspace}                 
 \def\Pc      {\ensuremath{\mathrm{c}}\xspace}

 \def\Pi      {\ensuremath{\mathrm{i}}\xspace}

 \def\Pp      {\ensuremath{\mathrm{p}}\xspace}

 \def\Ps      {\ensuremath{\mathrm{s}}\xspace}

 \def\thebaroffset{0.0em}
}
{

 \def\Ppi         {\ensuremath{\pi}\xspace}

 \def\Ppsi        {\ensuremath{\psi}\xspace}                 
                  
 \mathchardef\PDelta="7101
 \mathchardef\PXi="7104
 \mathchardef\PLambda="7103
 \mathchardef\PSigma="7106
 \mathchardef\POmega="710A
 \mathchardef\PUpsilon="7107
                  
 \def\PB      {\ensuremath{B}\xspace}                 
                  
 \def\PD      {\ensuremath{D}\xspace}

 \def\PJ      {\ensuremath{J}\xspace}                 
 \def\PK      {\ensuremath{K}\xspace}

 \def\Pb      {\ensuremath{b}\xspace}                 
 \def\Pc      {\ensuremath{c}\xspace}

 \def\Pi      {\ensuremath{i}\xspace}

 \def\Pp      {\ensuremath{p}\xspace}

 \def\Ps      {\ensuremath{s}\xspace}

 \def\thebaroffset{0.18em}
}
\newcommand{\offsetoverline}[2][\thebaroffset]{\kern #1\overline{\kern -#1 #2}}%

\makeatletter
\ifcase \@ptsize \relax
  \newcommand{\miniscule}{\@setfontsize\miniscule{4}{5}}
\or
  \newcommand{\miniscule}{\@setfontsize\miniscule{5}{6}}
\or
  \newcommand{\miniscule}{\@setfontsize\miniscule{5}{6}}
\fi
\makeatother

\DeclareRobustCommand{\optbar}[1]{\shortstack{{\miniscule (\rule[.5ex]{1.25em}{.18mm})}
  \\ [-.7ex] $#1$}}

\def\squark    {{\ensuremath{\Ps}}\xspace}
\def\squarkbar {{\ensuremath{\overline \squark}}\xspace}
\def\ssbar     {{\ensuremath{\squark\squarkbar}}\xspace}
\def\cquark    {{\ensuremath{\Pc}}\xspace}
\def\cquarkbar {{\ensuremath{\overline \cquark}}\xspace}
\def\ccbar     {{\ensuremath{\cquark\cquarkbar}}\xspace}
\def\bquark    {{\ensuremath{\Pb}}\xspace}

\def\pion   {{\ensuremath{\Ppi}}\xspace}

\def\pip    {{\ensuremath{\pion^+}}\xspace}
\def\pim    {{\ensuremath{\pion^-}}\xspace}

\def\kaon    {{\ensuremath{\PK}}\xspace}

\def\KorKbar {\kern \thebaroffset\optbar{\kern -\thebaroffset \PK}{}\xspace}

\def\Kp      {{\ensuremath{\kaon^+}}\xspace}
\def\Km      {{\ensuremath{\kaon^-}}\xspace}

\def\Kstarz  {{\ensuremath{\kaon^{*0}}}\xspace}

\def\Kstar   {{\ensuremath{\kaon^*}}\xspace}

\def\D       {{\ensuremath{\PD}}\xspace}

\def\DorDbar {\kern \thebaroffset\optbar{\kern -\thebaroffset \PD}\xspace}

\def\Dp      {{\ensuremath{\D^+}}\xspace}
\def\Dm      {{\ensuremath{\D^-}}\xspace}

\def\DpDm    {\ensuremath{\Dp {\kern -0.16em \Dm}}\xspace}

\def\B       {{\ensuremath{\PB}}\xspace}

\def\BorBbar {\kern \thebaroffset\optbar{\kern -\thebaroffset \PB}\xspace}

\def\Bd      {{\ensuremath{\B^0}}\xspace}

\def\BdorBdbar {\kern \thebaroffset\optbar{\kern -\thebaroffset \Bd}\xspace}

\def\Bs      {{\ensuremath{\B^0_\squark}}\xspace}

\def\BsorBsbar {\kern \thebaroffset\optbar{\kern -\thebaroffset \Bs}\xspace}

\def\Bds     {{\ensuremath{\B_{(\squark)}^0}}\xspace}

\def\jpsi     {{\ensuremath{{\PJ\mskip -3mu/\mskip -2mu\Ppsi}}}\xspace}

\def\Y#1S{\ensuremath{\PUpsilon{(#1S)}}\xspace}

\def\proton      {{\ensuremath{\Pp}}\xspace}
\def\antiproton  {{\ensuremath{\overline \proton}}\xspace}

\def\LorLbar     {\kern \thebaroffset\optbar{\kern -\thebaroffset \PLambda}\xspace}

\def\BF         {{\ensuremath{\mathcal{B}}}\xspace}

\newcommand{\decay}[2]{\ensuremath{#1\!\to #2}\xspace} 

\def\to                 {\ensuremath{\rightarrow}\xspace}

\def\AT#1     {\ensuremath{A_{\mathrm{T}}^{#1}}\xspace}           

\def\C#1      {\ensuremath{\mathcal{C}_{#1}}\xspace}                       
\def\Cp#1     {\ensuremath{\mathcal{C}_{#1}^{'}}\xspace}                    
\def\Ceff#1   {\ensuremath{\mathcal{C}_{#1}^{\mathrm{(eff)}}}\xspace}        
\def\Cpeff#1  {\ensuremath{\mathcal{C}_{#1}^{'\mathrm{(eff)}}}\xspace}       
\def\Ope#1    {\ensuremath{\mathcal{O}_{#1}}\xspace}                       
\def\Opep#1   {\ensuremath{\mathcal{O}_{#1}^{'}}\xspace}                    

\newcommand{\nospaceunit}[1]{\ensuremath{\text{#1}}}       
\newcommand{\aunit}[1]{\ensuremath{\text{\,#1}}}       

\newcommand{\tev}{\aunit{Te\kern -0.1em V}\xspace}
\newcommand{\gev}{\aunit{Ge\kern -0.1em V}\xspace}
\newcommand{\mev}{\aunit{Me\kern -0.1em V}\xspace}
\newcommand{\kev}{\aunit{ke\kern -0.1em V}\xspace}
\newcommand{\ev}{\aunit{e\kern -0.1em V}\xspace}
 
\newcommand{\mevc}{\ensuremath{\aunit{Me\kern -0.1em V\!/}c}\xspace}
\newcommand{\gevc}{\ensuremath{\aunit{Ge\kern -0.1em V\!/}c}\xspace}
\newcommand{\mevcc}{\ensuremath{\aunit{Me\kern -0.1em V\!/}c^2}\xspace}
\newcommand{\gevcc}{\ensuremath{\aunit{Ge\kern -0.1em V\!/}c^2}\xspace}

\def\mum  {\ensuremath{\,\upmu\nospaceunit{m}}\xspace}

\def\fb   {\ensuremath{\aunit{fb}}\xspace}
\def\invfb   {\ensuremath{\fb^{-1}}\xspace}

\def\ps   {\ensuremath{\aunit{ps}}\xspace}

\newcommand{\chisq}{\ensuremath{\chi^2}\xspace}

\newcommand{\chisqip}{\ensuremath{\chi^2_{\text{IP}}}\xspace}

\def\gsim{{~\raise.15em\hbox{$>$}\kern-.85em
          \lower.35em\hbox{$\sim$}~}\xspace}
\def\lsim{{~\raise.15em\hbox{$<$}\kern-.85em
          \lower.35em\hbox{$\sim$}~}\xspace}

\def\mrad{\aunit{mrad}\xspace}

\def\evtgen     {\mbox{\textsc{EvtGen}}\xspace}

\def\geant      {\mbox{\textsc{Geant4}}\xspace}

\def\photos     {\mbox{\textsc{Photos}}\xspace}

\def\pythia     {\mbox{\textsc{Pythia}}\xspace}

\def\tell1  {TELL1\xspace}
\def\ukl1   {UKL1\xspace}

\usepackage{cite} 
\usepackage{mciteplus}

\newcommand{\mygevc}{\ensuremath{{\mathrm{Ge\kern -0.1em V\!/}c}}\xspace}
\newcommand{\xx}{\ensuremath{\kern 0.5em }}

\newcommand{\BBspp}{\ensuremath{\Bds \to \proton \antiproton}\xspace}

\def\pp {\ensuremath{pp}\xspace}

\usepackage{multirow} 
\usepackage{booktabs} 
\usepackage{rotating}

\def\runI  {Run~1\xspace}
\def\runII {Run~2\xspace}

\def\Bds{{\ensuremath{\B^0_{\kern -0.1em{\scriptscriptstyle (}\kern -0.05em\squark\kern -0.03em{\scriptscriptstyle )}}}}\xspace}

\newcommand{\PPbar}{\proton \antiproton}

\newcommand{\PPbarKK}{\proton \antiproton \Kp \Km}
\newcommand{\PPbarKPi}{\proton \antiproton \Kp \pim}
\newcommand{\PPbarPPbar}{\proton \antiproton \proton \antiproton}

\def\PPbarHH{\ensuremath{\proton\antiproton h h^{\prime}}\xspace}

\def\BdJPsiKst{\decay{\Bd}{\jpsi\Kstarz}}

\def\BdJPsiKstDetailed{\decay{\Bd}{{\jpsi}(\to{\proton\antiproton}){\Kstarz}(\to{\Kp\pim})}}

\newcommand{\BdPiPi}{\texorpdfstring{\decay{\Bd}{\pip \pim}}{}}

\newcommand{\BdPPbar}{\texorpdfstring{\decay{\Bd}{\proton \antiproton}}{}}
\newcommand{\BPPbar}{\texorpdfstring{\decay{\Bds}{\proton \antiproton}}{}}

\def\BdPPbarHH{\decay{\Bd}{\proton\antiproton h h^{\prime}\xspace}}
\newcommand{\BdPPbarPPbar}{\texorpdfstring{\decay{\Bd}{\proton  \antiproton \proton \antiproton}}{}}
\newcommand{\BsPPbarPPbar}{\texorpdfstring{\decay{\Bs}{\proton  \antiproton \proton \antiproton}}{}}
\newcommand{\BPPbarPPbar}{\texorpdfstring{\decay{\Bds}{\proton  \antiproton \proton \antiproton}}{}}

\newcommand{\BdPJpsiKstar}{\texorpdfstring{\decay{\Bd}{\jpsi \Kstarz}}{}}

\newcommand{\BsJPsiPhi}{\texorpdfstring{\decay{\Bs}{\jpsi \phi}}{}}
\newcommand{\BsJPsiPhiDetailled}{\texorpdfstring{\decay{\Bs}{\jpsi(\to{\proton \antiproton}) \phi(\to{\Kp\Km})}}{}}

\newcommand{\BsPPbar}{\texorpdfstring{\decay{\Bs}{\proton \antiproton}}{}}

\newcommand{\BsJpsiPhi}{\texorpdfstring{\decay{\Bs}{\jpsi \phi}}{}}

 \def\mppKpi     {\ensuremath{m_{\PPbarKPi}}\xspace}                 
 \def\mpp        {\ensuremath{m_{\PPbar}}\xspace}                 
 \def\mKpi       {\ensuremath{m_{\Kp\pim}}\xspace}                 
 \def\mppKK      {\ensuremath{m_{\PPbarKK}}\xspace}                 
 \def\mKK        {\ensuremath{m_{\Kp\Km}}\xspace}

\def\prodProbNNp      {\ensuremath{\prod_i{\cal P}_i(\proton)}\xspace}    
\usepackage{longtable} 
\usepackage{subcaption}

\begin{document}

\renewcommand{\thefootnote}{\fnsymbol{footnote}}
\setcounter{footnote}{1}

\begin{titlepage}
\pagenumbering{roman}

\vspace*{-1.5cm}
\centerline{\large EUROPEAN ORGANIZATION FOR NUCLEAR RESEARCH (CERN)}
\vspace*{1.5cm}
\noindent
\begin{tabular*}{\linewidth}{lc@{\extracolsep{\fill}}r@{\extracolsep{0pt}}}
\ifthenelse{\boolean{pdflatex}}
{\vspace*{-1.5cm}\mbox{\!\!\!\includegraphics[width=.14\textwidth]{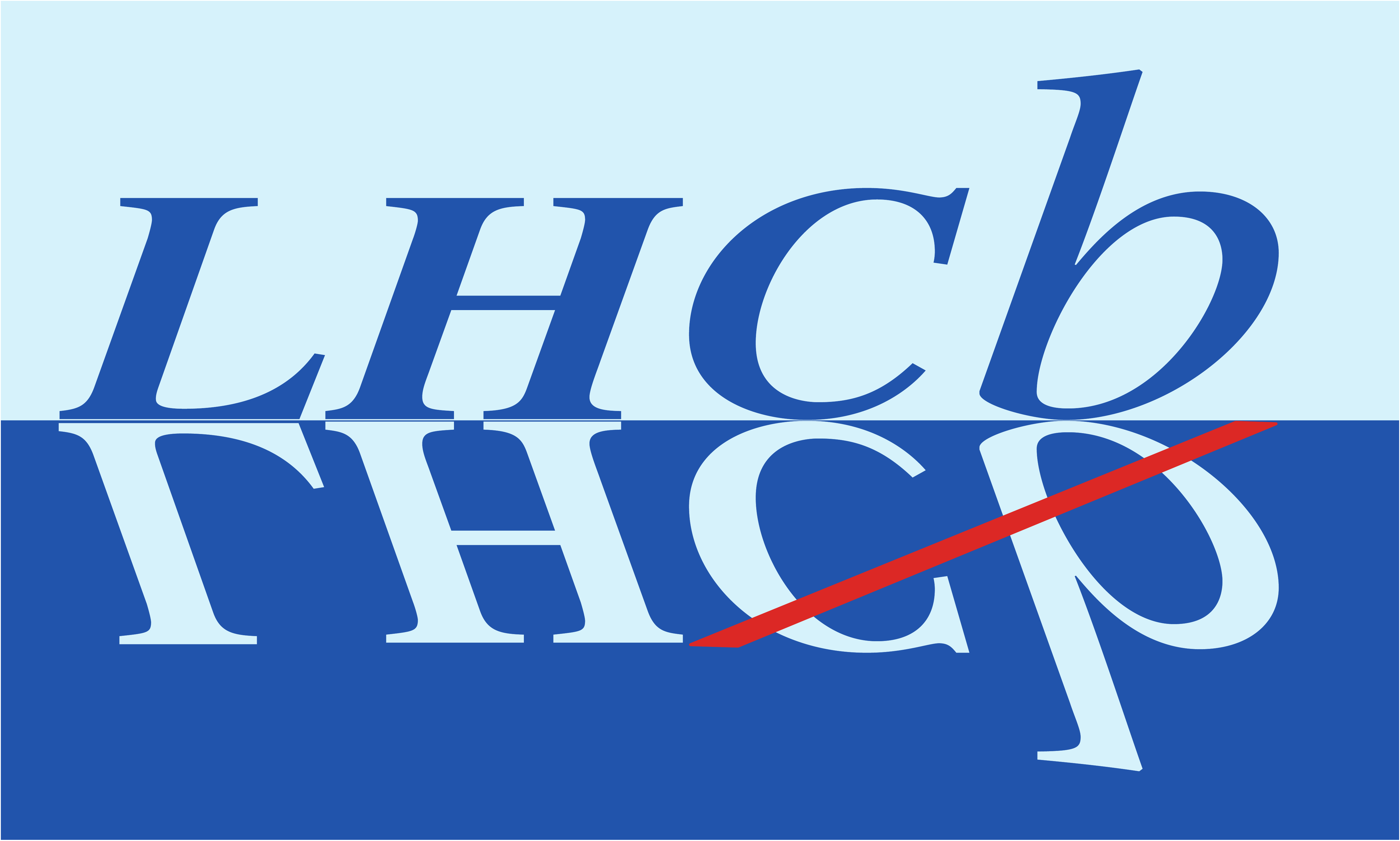}} & &}
{\vspace*{-1.2cm}\mbox{\!\!\!\includegraphics[width=.12\textwidth]{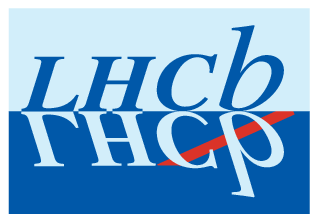}} & &}
\\
 & & CERN-EP-2022-206 \\
 & & LHCb-PAPER-2022-032 \\  
 & & 15 November 2022 \\ 
 & & \\
\end{tabular*}

\vspace*{2.5cm}

{\normalfont\bfseries\boldmath\huge
\begin{center}
  \papertitle 
\end{center}
}

\vspace*{2.0cm}

\begin{center}
\paperauthors
 \end{center}

\vspace{\fill}

\begin{abstract}
  \noindent
Searches for the rare hadronic decays \BdPPbarPPbar and \BsPPbarPPbar are performed using proton-proton collision data recorded by the \lhcb experiment and corresponding to an integrated luminosity of 9\invfb.
Significances of $9.3\,\sigma$ and $4.0\,\sigma$, including statistical and systematic uncertainties, are obtained for the \BdPPbarPPbar and \mbox{\BsPPbarPPbar} signals, respectively.
The branching fractions are measured relative to the topologically similar normalisation decays \mbox{\BdJPsiKstDetailed} and \BsJPsiPhiDetailled.
The branching fractions are measured to be
\mbox{$ \BF(\BdPPbarPPbar) = ( 2.2 \pm 0.4 \pm 0.1 \pm 0.1 ) \times 10^{-8} $}
and
\mbox{$ \BF(\BsPPbarPPbar) = ( 2.3 \pm 1.0 \pm 0.2 \pm 0.1 ) \times 10^{-8}$}.
In these measurements, the first uncertainty is statistical, the second is systematic and the third one is due to the external branching fraction of the normalisation channel.

\end{abstract}

\vspace*{2.0cm}

\begin{center}
  Submitted to Phys.~Rev.~Lett.
\end{center}

\vspace{\fill}

{\footnotesize 
\centerline{\copyright~\papercopyright. \href{\paperlicenceurl}{\paperlicence}.}}
\vspace*{2mm}

\end{titlepage}

\newpage
\setcounter{page}{2}
\mbox{~}

\renewcommand{\thefootnote}{\arabic{footnote}}
\setcounter{footnote}{0}

\cleardoublepage

\pagestyle{plain}
\setcounter{page}{1}
\pagenumbering{arabic}

The physics of \B mesons decays to charmless, hadronic final states with \PPbar pairs is not well understood.
Non-perturbative strong interactions combine with short-distance weak decays to produce final states that display non-trivial patterns.
Differences in the way \B mesons decay to baryonic versus purely mesonic final states have been found since the first experimental measurement of a baryonic \B decay was reported by the CLEO collaboration~\cite{Fu:1996qt}.
The branching fraction \mbox{$\BF(\BdPPbar) = (1.3 \pm 0.2 ) \times 10^{-8}$}~\cite{LHCb-PAPER-2022-004} is two orders of magnitude lower than $\BF(\BdPiPi) = (5.1 \pm 0.2 ) \times 10^{-6}$~\cite{PDG2022}, while four-body \BdPPbarHH decays are not as suppressed relative to the corresponding two-body decays $ \Bd \to h h^{\prime } $ where $ h h^{\prime } $ denotes a meson pair.
For \mbox{$h h^{ \prime }$ = $\Kp\pim$}, the ratio of four-body to two-body branching fractions is about $0.43 \pm 0.03$, while for \mbox{$ h h^{ \prime } $ = $ \pip \pim$}, the ratio is about $ 0.57 \pm 0.04 $~\cite{PDG2022}.
The knowledge of related \Bs decays is more limited.
The branching fractions for the $ p \bar p K^+ K^- $, $ p \bar p K^+ \pi^- $, and $ p \bar p \pi^+ \pi^- $ final states are $ (4.5 \pm 0.5 ) \times 10^{-6} $, $ ( 1.4 \pm 0.3 )  \times 10^{-6} $, and $ ( 4.3 \pm 2.0 ) \times 10^{-7} $, respectively,~\cite{LHCb-PAPER-2017-005}.
The decay \BsPPbar has not been observed; the LHCb collaboration recently reported an upper limit of $\BF(\BsPPbar) < 4.4 \, (5.1) \times 10^{-9}$ at the 90\% (95\%) confidence level (CL)~\cite{LHCb-PAPER-2022-004}.
The branching fractions of multi-body baryonic decay modes may be significantly increased due to a threshold enhancement effect in the baryon-antibaryon invariant mass spectrum~\cite{Hou:2000bz, Bevan:2014iga, Hsiao:2019wyd,Huang:2021qld}, while two-body baryonic decays (such as \BBspp) are suppressed~\cite{Hsiao:2014zza, Cheng:2014qxa, Huang:2021qld}.

The fully baryonic decay \BdPPbarPPbar has not been observed previously; the \babar collaboration reports an upper limit  of $2 \times 10^{-7}$ at 90\% CL~\cite{BABAR:2018erd}.
No prior search for \BsPPbarPPbar has been reported. 
While no theoretical prediction related to the purely baryonic \BPPbarPPbar decays is currently available, a first measurement of the corresponding branching fractions would allow to better understand the underlying dynamics.
The goal of the analysis presented here is to probe \BPPbarPPbar branching fractions with sensitivities of ${\cal O} (10^{-8} )$, similar to that for \BPPbar and an order of magnitude lower than the \BdPPbarPPbar limit from the \babar collaboration.

We search for \BPPbarPPbar using the full \runI and \runII proton-proton collision data sets.\footnote{Charge-conjugate decays are implied throughout this article, unless explicitly stated otherwise.}
These were collected by the \lhcb experiment over the period 2011--2018 at center-of-mass energies of 7, 8, and 13\tev with integrated luminosities of approximately 1, 2, and 6\invfb, respectively.
A blind selection procedure is used, with unblinding performed in two steps: first the \runI then the \runII\ \PPbarPPbar data. 
The \BdPPbarPPbar and \BsPPbarPPbar\ yields are determined from fits to the integrated data sets with different selection criteria.
Anticipating the \Bs signal to be much smaller than the \Bd signal, tighter selection criteria for the \Bs study are used to further suppress background.

The branching fractions of the signal decays are measured simultaneously across all data-taking periods relative to the topologically similar \BdJPsiKstDetailed and \BsJPsiPhiDetailled normalisation channels using
\begin{eqnarray}
\label{eq:BF_B24P}
\BF(\BdPPbarPPbar) & = & \BF_{\textrm{vis}}(\BdPJpsiKstar) \times \frac{N(\BdPPbarPPbar)}{N(\BdPJpsiKstar)}\times\frac{\epsilon(\BdPJpsiKstar)}{\epsilon(\BdPPbarPPbar)}\,,\\
\label{eq:BF_Bs24P}
\BF(\BsPPbarPPbar) & = & \BF_{\textrm{vis}}(\BsJpsiPhi) \times \frac{N(\BsPPbarPPbar)}{N(\BsJpsiPhi)} \times \frac{\epsilon(\BsJpsiPhi)}{\epsilon(\BsPPbarPPbar)}\,,
\end{eqnarray}
where $N$ denotes a measured yield and $\epsilon$ denotes a combination of geometrical acceptance of the \lhcb detector and reconstruction, trigger and selection efficiencies. 
The symbol $\BF_{\textrm{vis}}$ denotes the  branching fraction of the nominal final state multiplied by the branching fractions of its resonances to their decay products.
The normalisation channels, \mbox{\BdJPsiKst} and \mbox{\BsJPsiPhi}, are chosen because their branching fractions are well measured: 
$ (1.27 \pm 0.05) \times 10^{-3} $ 
and 
$ (1.04 \pm 0.04 ) \times 10^{-3}$, respectively~\cite{PDG2022}. 
Simultaneously fitting the ratios of signal \BPPbarPPbar rates to normalisation channel rates mitigates many of the possible systematic uncertainties.

The \lhcb detector~\cite{LHCb-DP-2008-001,LHCb-DP-2014-002} is a single-arm forward spectrometer covering the pseudorapidity range $2<\eta <5$, designed for the study of particles containing \bquark or \cquark quarks.
The detector includes a high-precision tracking system consisting of a silicon-strip vertex detector surrounding the \pp interaction region~\cite{LHCb-DP-2014-001}, a large-area silicon-strip detector located upstream of a dipole magnet with a bending power of about $4{\mathrm{\,Tm}}$, and three stations of silicon-strip detectors and straw drift tubes~\cite{LHCb-DP-2017-001} placed downstream of the magnet.
The tracking system provides the measurements of the track momentum and impact parameter (IP) and is used to reconstruct primary vertices (PVs). 
Different types of charged hadrons are distinguished using particle identification (PID) information from two ring-imaging Cherenkov detectors~\cite{LHCb-DP-2012-003}.
The online event selection is performed by a trigger~\cite{LHCb-DP-2012-004}, which consists of a hardware stage followed by a software stage, comprising two levels, which applies a full event reconstruction.
All candidates must satisfy at least one of two criteria --- either a decision to accept the event was completely {\em independent} of the tracks used to form the candidate (TIS) or a positive decision was dependent {\em on} the tracks in the signal and no others (TOS).

Simulated samples are used to study the properties of the signal, normalisation and background channels. 
Proton-proton collisions are generated by \pythia~\cite{Sjostrand:2007gs} with a specific \lhcb configuration~\cite{LHCb-PROC-2010-056}.
Decays of unstable particles are described by \evtgen~\cite{Lange:2001uf}, in which final-state radiation is generated using \photos~\cite{Photos-2}.
The interactions of the generated particles with the detector material, and their responses, are implemented using the \geant toolkit~\cite{Agostinelli:2002hh,Allison:2006ve}, as described in Ref.~\cite{LHCb-PROC-2011-006}.

In the offline selection, \B candidates are formed from tracks required to satisfy quality criteria, and  to have a momentum in the range $10 < p < 110\gevc$.
Individual track trajectories must be inconsistent with originating at a PV using  the criterion $\chisqip > 25$, where \chisqip is the difference between the vertex-fit \chisq of a PV reconstructed with and without the track in question.
The distance of closest approach between any two tracks of a \B candidate is required to be less than 300\mum.
The \B candidate momentum is required to point back to its PV by requiring $\chisqip < 25$.
The angle between the candidate momentum vector and the line connecting the associated PV and the decay vertex of the candidate must be less than 14\mrad.
The \B vertex is required to be well separated from the PV and the \B candidate must have a decay time greater than 1.0\ps.
For the \PPbarHH data samples additional requirements are made on the final state invariant masses around the known masses of the \jpsi, \Kstarz or $\phi$ mesons: $|m_{\PPbar}-3097 | < 60 \mevcc$, $|m_{\Kp\pim}-895.5 | < 200 \mevcc$ and $|m_{\Kp\Km}-1019.5 | < 30 \mevcc$.

The two most powerful discriminating variables between signal and background are the $\chisqip$ of the \Bds candidate with respect to the PV and the quantity \prodProbNNp.
The latter is defined as the product of the probabilities of the four protons being correctly identified.
Since the PID algorithms were tuned differently for \runI and \runII, \prodProbNNp is optimized separately for the two running periods.
Using \PPbarPPbar data in the sideband mass ranges $m(\PPbarPPbar) \in (4950 - 5240) \cup (5407 - 5650)\mevcc$ and simulated \Bd and \Bs signals, the selections applied to these variables are chosen to maximize ${\rm S}/\sqrt{\rm B}$, where S and B are the signal and  background yields in the signal region $m(\PPbarPPbar) \in [5240, 5407]\mevcc$, with the constraint that $ {\rm S}/\sqrt{{\rm S}+{\rm B}} $ is greater than 95\% of its maximum value.

The expected yields from \BdPPbarPPbar decays are larger than the corresponding \Bs decay.
At tree level, \BdPPbarPPbar decays occur mainly via a $W$-emission diagram, while \BsPPbarPPbar decays occur through Cabibbo suppressed $W$-exchange diagrams. 
The \Bs production rate is further suppressed by $f_s/f_d\sim 25\%$\cite{LHCb:2021qbv} compared to that of the \Bd. 
Thus, different working points are chosen: \textit{tight} and \textit{very tight} selections for the \Bd and \Bs searches, respectively.

Expected signal rates in \runI are estimated from a previous study based on \runI\ \PPbarPPbar data~\cite{WamplerThesis}, while those expected in \runII are extrapolated from the \runI measured \BdPPbarPPbar yield.
Prior to unblinding the \runI data, a list of tight and very tight \prodProbNNp working points for the \runII data as a function of a large range of possible central values for the \Bd signal in \runI were established.
In choosing the working points, only \prodProbNNp values that are integer multiples of 0.05 in the range $(0.10 - 0.90)$ were considered to avoid possible biases associated with fine tuning.
The \prodProbNNp requirement for the very tight selection increases the background rejection by a factor of four while reducing the signal efficiency by $\sim$40\% with respect to the tight selection.
For the $\chisqip$ variable, a common requirement $\chisqip<1.8$ is used for both channels in all running periods.
Signal and normalisation efficiencies are estimated from simulation weighted to match the data distributions of the $\chisqip$ of the \Bds candidate, the track multiplicity in the event and the momentum and pseudorapidity of each daughter particle.
The weights are obtained from background-subtracted normalisation data samples with loose requirements applied (all selection criteria except those on $\Bds \ \chisqip$ and \prodProbNNp).

\begin{figure}[tb]
    \centering
    \includegraphics[width=0.48\textwidth]{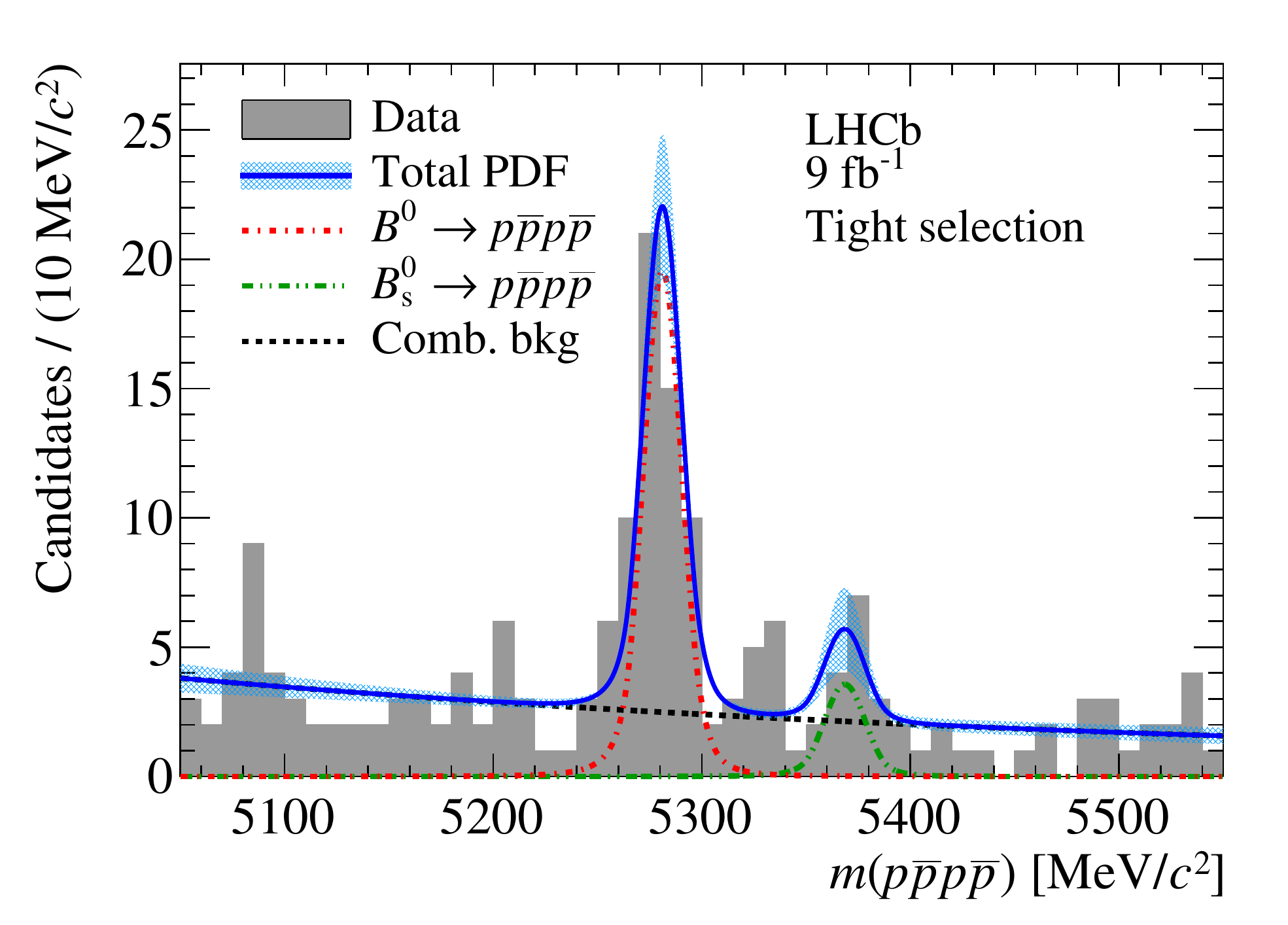}
    \includegraphics[width=0.48\textwidth]{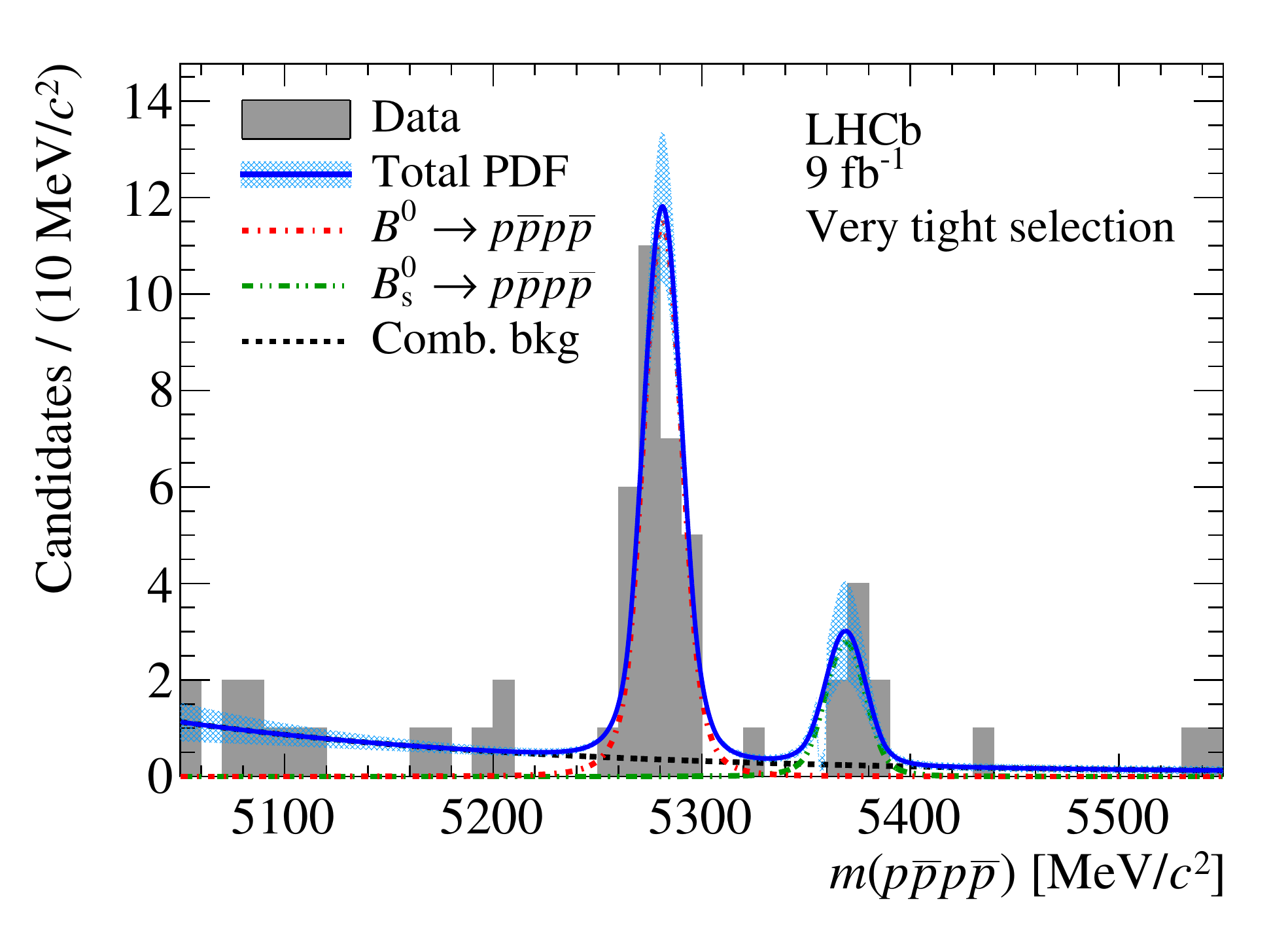}
    \caption{Combined \runI plus \runII invariant-mass distributions of  
    \BPPbarPPbar candidates satisfying (left) the tight
    and (right) the very tight selection criteria discussed in the text.
    The fit results (solid blue lines) for these samples are
     shown together with the fit model components. 
     The hashed cyan band corresponds to the $1\,\sigma$ model uncertainty based on the fit covariance matrix. }    
    \label{fig:SignalMasses}
\end{figure}

The invariant mass ($m(\PPbar\PPbar)$) distributions of the candidates satisfying the tight and very tight selection criteria are shown in Fig.~\ref{fig:SignalMasses}.
The candidates selected with the tight (very tight) cuts are used to measure $\BF(\BdPPbarPPbar)$ $(\BF(\BsPPbarPPbar ))$.
The fits shown here are used only to determine the statistical significances of the signals.
The (common) signal shapes are assumed to be described by double-sided Crystal Ball (DSCB) functions~\cite{Skwarnicki:1986xj} with tail parameters fixed to those found in simulation.
The mean of the \Bd mass distribution is Gaussian-constrained to the value measured for the normalisation channel \BdJPsiKst and the mean of the \Bs mass distribution is fixed to that of the \Bd plus the known $ \Bs - \Bd $ mass difference~\cite{PDG2022}.
The signal widths are fixed to the values measured for the corresponding normalisation channels scaled by the ratio of signal to normalisation widths measured in the simulation.
The background shapes are assumed to be exponential with coefficients that are left free to vary in the fit.
The \Bd and \Bs signal yields obtained from the fit to the samples with tight and very tight selection criteria are $ 48.2 \pm 8.1 $ and $ 7.1 \pm 2.9 $, respectively.
The significances of the signal yields are computed from likelihood scans using Wilks' theorem~\cite{Wilks:1938dza}.
They are found to be $ 9.3 \, \sigma $ and $ 4.0 \, \sigma $, including statistical uncertainties only, for the \BdPPbarPPbar and \BsPPbarPPbar signal respectively.

For each channel, the absolute branching fraction is measured from a simultaneous unbinned maximum-likelihood fit to the signal and normalisation channels in which the corresponding yields are free to vary (see Eq.~\ref{eq:BF_B24P}).
The data are divided into four running periods during which operating conditions varied (years 2011-2012, 2015-2016, 2017, and 2018); these data are fitted simultaneously.
The signal shapes described above are also used in this fit.
Common means and widths are used to describe the \Bd and \Bs mass distributions between the signal and normalisation channels.
The \Bd and \Bs mass and width parameters vary freely in the fit; they are common to the signal and normalisation modes but the signal widths are scaled by the ratio of signal to normalisation widths in the simulation.
Several shape parameters of the normalisation channels are common between the three data-taking periods in \runII.
The efficiencies for both signal and normalisation channels are determined separately for each running period.

For each normalisation channel, \mbox{\BdJPsiKstDetailed} and \mbox{\BsJPsiPhiDetailled}, the signal model for the corresponding \PPbarHH sample is modeled as a three-dimensional (3D) probability density function in $m_{\PPbarHH}$, \mpp, and $m_{h h^{\prime}}$.
The \BdJPsiKst (\BsJPsiPhi) contributions to the \mppKpi (\mppKK) and \mpp spectra are parameterised using DSCB functions to describe the \Bd (\Bs) and \jpsi shapes.
The \jpsi resonance is very narrow, so its shape is completely dominated by resolution effects.
The $\Kstar(892)^0$ shape in the \mKpi spectrum is modeled using a relativistic Breit--Wigner (RBW) function.
As its natural width is so large, resolution effects are negligible.
Since the natural width of the $\phi$ meson, $\Gamma = (4.23 \pm 0.01) \mevcc$~\cite{PDG2022}, is comparable to the expected resolution ($\sim$10\mevcc) obtained from simulation, the \BsJPsiPhi contribution to the \mKK spectrum is parameterised using a RBW function convolved with a Gaussian function accounting for resolution effects.
An $S$-wave component, denoted $(K\pi)_0$, is modeled in the \mKpi spectrum distribution using a LASS parametrization that describes non-resonant and $K^*_0(1430)^0$ $S$-wave contributions~\cite{Aston:1987ir,BaBar:2008lan}; this component is modeled in the \PPbarKPi and \PPbar invariant mass distributions with the same shape as the \BdJPsiKst component.
The invariant mass distributions of the normalisation channels, with the 3D fit projections superimposed from which the \jpsi\Kstarz and $\jpsi\phi$ normalisation yields are extracted, are given in the Supplemental Material~\cite{SuppMat}.

Sources of systematic uncertainties from the fit models, the efficiencies as estimated from simulation, particle identification and the branching fractions of the normalisation channels are considered.
A summary is presented in Table~\ref{tab:summary_syst_BF}.
A subset of the systematic uncertainties affects the statistical significances of the \BPPbarPPbar yields.
The signal shapes are varied in two ways: (i) the DSCB tail parameters are varied and (ii) the DSCB function is replaced with a Johnson SU distribution~\cite{JohnsonSu}.
The exponential background shapes are replaced with linear shapes.
The largest change in the \Bd (\Bs) yield is $ \sim $1\% (9\%) which we assign as the associated systematic uncertainty.
In both cases, the absolute uncertainty in signal yield is $ \sim $0.5 events, which is negligible compared to the corresponding statistical uncertainties.

Uncertainties on the material budget and particle interaction cross-sections lead to uncertainties on the efficiency ratios. 
These are estimated to be $ 0.8\% $ and $ 1.0\% $ for $ \BF(\BdPPbarPPbar ) $ and $ \BF(\BsPPbarPPbar ) $, respectively.
Differences between PID efficiencies in data and simulation are corrected using a combination of simulated and data calibration samples.
The corrections depend on both the sample size of the simulation samples and the kernel densities used to weight the data calibration samples to match the kinematic distributions of the signals.
The overall uncertainties from PID are estimated to be $ \sim $1.5\% in both cases.
Two sources of systematic uncertainties associated with the efficiencies estimated from the simulation are determined.
First, the uncertainties on the nominal efficiencies due to the size of the simulated samples are evaluated by applying a Gaussian constraint in the fit. 
Second, the uncertainties from the weights used to correct discrepancies between simulation and data are conservatively estimated from an alternative fit with unweighted efficiencies.
The corresponding systematic uncertainties on both signal branching fractions are estimated to be $ \sim $1\% and $ \sim $3\% for these two sources, respectively.
The latter covers uncertainties, of typically $\sim $1\%, due to imperfect description of the detector in the simulation for relative hardware trigger efficiencies between two topologically similar decay modes~\cite{LHCb-PAPER-2022-004}.
The uncertainties associated with the normalisation channel branching fractions are taken from~\cite{PDG2022}.

\begin{table}[t]
\caption{Summary of systematic uncertainties on $\BF( \BPPbarPPbar )$ relative to the statistical uncertainties in [\%]. The total systematic uncertainty is calculated as the quadratic sum of the individual systematic uncertainties. 
\label{tab:summary_syst_BF}}
\begin{center}
\begin{tabular}{l c c }
\toprule 
Systematic source   & $\BF(\BdPPbarPPbar)$ & $\BF(\BsPPbarPPbar)$ \\ 
\midrule 
Efficiencies (sample size) 	&  $\phantom{1}5$ & $\phantom{1}3$ \\ 
Efficiencies (weights) 		&            $16$ & $\phantom{1}9$ \\ 
PID 						&  $\phantom{1}8$ & $\phantom{1}3$ \\ 
Tracking 					&  $\phantom{1}5$ & $\phantom{1}2$ \\ 
Fixed PDF parameters 		&  $\phantom{1}5$ & $\phantom{1}2$ \\ 
Signal model 				&  $\phantom{1}1$ & $\phantom{1}4$ \\ 
Background model 			&  $\phantom{1}8$ &           $18$ \\ 
\midrule 
Total systematic 			&            $22$ &           $21$ \\ 
\midrule 
Normalisation $\BF$ 		&            $24$ &           $13$ \\ 
\bottomrule 
\end{tabular} 
\end{center} 
\end{table}

Finally, potential remaining sources of systematic uncertainties are studied by dividing the data into disjoint samples and comparing the branching fractions measured in each subset. 
Five sets of disjoint subsamples are defined, with candidates separated according to: the two magnet polarities, the four running periods, four bins of candidate $ B $ momentum, four bins of candidate $ B $ pseudorapidity, and three mutually exclusive hardware trigger categories (TIS only, TOS only, and the overlap between them).
For each set of subsamples, $ i $, a $ p $-value ($ p_i $) is determined for the hypothesis that the observed variations are consistent with statistical fluctuations using a $ \chi^2 $ test comparing the nominal branching fractions with those observed in the subsamples.
The ensemble $ p $-value defined as the product $ \prod_i p_i $ is then computed.
Ensemble $ p $-values of 0.15 and 0.55 are observed for the $\BF(\BdPPbarPPbar)$ and $\BF(\BsPPbarPPbar)$ measurements, respectively, and no additional systematic uncertainties are included.

Using the tight and very tight selections, the \BdPPbarPPbar and \BsPPbarPPbar branching fractions are measured to be \mbox{$\BF(\BdPPbarPPbar) = (2.2 \pm 0.4 \pm 0.1 \pm 0.1) \times 10^{-8}$} and \mbox{$ \BF(\BsPPbarPPbar) = (2.3 \pm 1.0 \pm 0.2 \pm 0.1 ) \times 10^{-8} $}, where the first uncertainty is statistical, the second is systematic and the third is due to the external branching fraction of the normalisation channel.
The significances that the \BdPPbarPPbar and \BsPPbarPPbar signals differ from zero, accounting for systematic uncertainties associated with signal and background shapes, are $ 9.3 \, \sigma $ and $ 4.0 \, \sigma $, respectively.
The branching fraction ratios are $\BF(\BdPPbarPPbar)/\BF(\BdJPsiKstDetailed) = (1.24 \pm 0.21 \pm 0.04)\times 10^{-2}$ and $\BF(\BsPPbarPPbar)/\BF(\BsJPsiPhiDetailled) = (2.1 \pm 0.9 \pm 0.2)\times 10^{-2}$, where the first uncertainty is statistical and the second is systematic.

The sizes of the \PPbarPPbar samples are too limited to quantitatively study possible \ccbar contributions that might be produced by tree-level, CKM-favored amplitudes, such as $\Bds \to \jpsi(\to \PPbar) \PPbar$.
By excluding \PPbarPPbar candidates if any \PPbar invariant mass is greater than $2850 \mevcc$, as shown in the Supplemental Material~\cite{SuppMat}, the data clearly demonstrate the presence of a dominant charmless contribution.
With the \ccbar veto the efficiencies for pure phase-space decays (ignoring anti-symmetrisation of amplitudes for fermion pairs) are reduced by 40--50\% according to the simulation. 
The \BdPPbarPPbar and \BsPPbarPPbar branching fractions measured with this additional requirement are $( 1.6 \pm 0.4 ) \times 10^{-8} $ and $ (2.2 \pm 1.2 ) \times 10^{-8} $ (statistical uncertainties only), respectively, consistent with those measured over the full phase space.
The \BdPPbarPPbar and \BsPPbarPPbar significances, with the \ccbar vetoes applied, are  $6.5\,\sigma$ and $3.6\,\sigma$.
A qualitative examination of the $m_{p \bar p}$ spectra in the \Bd sample is consistent with ${\cal B} ( \Bd \to J/\psi p \bar p) \times {\cal B} (J/\psi \to p \bar p )$ at the expected level of ${\cal O} (10^{-9})$~\cite{PDG2022} and no other obvious resonant contributions.
These observations support the hypothesis that both the \BdPPbarPPbar and the \BsPPbarPPbar decays proceed primarily through charmless transitions.

In summary, we have searched for the decays \BdPPbarPPbar and \BsPPbarPPbar using the full \lhcb \runI and \runII data sets and report branching fractions for both.
Significances of $9.3\,\sigma$ and $4.0\,\sigma$, including statistical and systematic uncertainties, are measured for \BdPPbarPPbar and \BsPPbarPPbar signals, respectively.
We observe that $\BF(\BdPPbarPPbar)$ is about an order of magnitude lower than the upper limit previously reported by the \babar collaboration.
This branching fraction is about twice that of $\BdPPbar$, which is in contrast to the \BdPPbarHH modes which have smaller branching fractions than the $\Bd \to h h^{\prime}$ channels.

No significant evidence for resonant substructure associated with either decay is observed.
The data suggest that $\BF(\BsPPbarPPbar)$ is of the same order of magnitude as $\BF(\BdPPbarPPbar)$, an unanticipated result.
A branching fraction as large as ${\cal O} (10^{-8})$ for the \Bs decay is difficult to explain simply as a tree-level process since this decay requires both CKM-suppressed production of an $ \bar s $ quark and strong scattering of the short-distance $ \ssbar $ system to a non-strange or hidden strangeness final state.
The patterns of \B-meson decays to final states with baryon pairs probe non-perturbative QCD and are complementary to the patterns of purely mesonic decays.
The data sets anticipated from Run 3, with a factor of 10 greater dataset than \runII, should be large enough to study the amplitude structures of \BdPPbarPPbar decays and definitively confirm or disprove the large \BsPPbarPPbar branching fraction suggested by the current data.

\clearpage
\addcontentsline{toc}{section}{References}
\setboolean{inbibliography}{true}
\bibliographystyle{LHCb}
\bibliography{main,LHCb-PAPER,LHCb-CONF,LHCb-DP,LHCb-TDR,standard,MyBib}

\end{document}